\documentclass{article}

\usepackage{arxiv}

\usepackage[numbers]{natbib}

\usepackage{algorithm}
\usepackage{algorithmic}
\usepackage{hyperref}
\usepackage{url}
\usepackage{mathtools}
\usepackage{booktabs} 
\usepackage{makecell}
\usepackage{xcolor}
\usepackage[capitalize,noabbrev]{cleveref}

\usepackage{amsmath,amssymb,amsfonts,amsbsy,amsthm}
\usepackage{graphicx}
\usepackage{textcomp}
\usepackage{xcolor}
\usepackage{bm}
\usepackage{caption}
\usepackage{subcaption}
\usepackage{multirow}
\usepackage{latexsym}
\usepackage{ifthen}
\usepackage{tabularx}
\usepackage[normalem]{ulem}
\usepackage{xspace}
\usepackage{paralist}	
\usepackage{breqn}
\usepackage{color}
\usepackage[utf8]{inputenc}
\usepackage{arydshln}

\newtheorem{definition}{Definition}

\newcommand{\bx}{{\bm x}}
\newcommand{\bp}{{\bm P}}
\newcommand{\bk}{\mathcal{K}} 
\newcommand{\ed}{\bm \epsilon}

\newcommand{\bv}{{\bm v}}
\newcommand{\bz}{{\bm z}}

\newcommand{\by}{{\bm y}}  
  
\newcommand{\ad}{{\mathcal{A}}}  

\newcommand{\on}{\Omega}  
\newcommand{\cn}{\overline{\Omega}}  
\newcommand{\btheta}{{\bm \theta}}
\newcommand{\xc}{{\bm x}^{\star}}
\newcommand{\yc}{{\bm y}^{\star}}
\newcommand{\ec}{{\bm \epsilon}^{\star}}
\newcommand{\kc}{{\mathcal{K}}^{\star}}
\newcommand{\pc}{{\bm P}^{\star}}

\title{An alternative approach for distributed parameter estimation under Gaussian settings}

\author{ Subhro Das \\
MIT-IBM Watson AI Lab, IBM Research\\
Cambridge, MA, USA \\
\texttt{subhro.das@ibm.com}
}

\date{}
\renewcommand{\headeright}{}
\renewcommand{\undertitle}{}

\begin{document}

\maketitle

\begin{abstract}
This paper takes a different approach for the distributed linear parameter estimation over a multi-agent network. The parameter vector is considered to be stochastic with a Gaussian distribution. The sensor measurements at each agent are linear and corrupted with additive white Gaussian noise. Under such settings, this paper presents a novel distributed estimation algorithm that fuses the the concepts of consensus and innovations by incorporating the consensus terms (of neighboring estimates) into the innovation terms. Under the assumption of distributed parameter observability, introduced in this paper, we design the optimal gain matrices such that the distributed estimates are consistent and achieves fast convergence. 
\end{abstract}


\section{Introduction}
Distributed parameter estimation has been studied significantly over the last two decades~\cite{chong2017forty,ding2019survey,ge2019distributed} for application in cyber-physical systems and wireless sensor networks. In distributed inference, there exists a class of \emph{consensus+innovations} distributed estimation algorithms that generalize classical distributed consensus by combining cooperation among agents (consensus) with assimilation of their observations (innovations)~\cite{kar2012distributed,karmoura-spm2013,kar2011convergence}. 
The problem is to estimate an underlying parameter vector by a large-scale network of inexpensive and low-power sensors, such that the measurements are sparse and corrupted by noise. 

Here we point out that distributed estimation framework is fundamentally different from federated learning~\cite{konevcny2016federated}, where a federated (fusion) server exists that performs some sort of aggregation (averaging in most cases) of the local parameters. However, in the distributed setting the data is distributed across several agents and there is no fusion center. Each agent computes their own local estimates and shares those estimates with their neighboring agents in the network.

Although the primary focus has been on deterministic parameter~\cite{ribeiro2004distributed,stankovic2010decentralized}, in this paper we consider a stochastic parameter that follows a Gaussian distribution. 
In this paper, we introduce a new definition of distributed parameter observability which is a weaker criteria than the distributed observability conditions considered in the literature. The key contribution of this work is the novel distributed algorithm that leverages the estimates from its neighbors and performs consensus within the innovation process. 
Such fusion is different from the consensus on pseudo-innovations in~\cite{das2013Allerton,das2013Asilomar}. 
This  technique facilitates the design of the optimal gain matrices, in contrast to scalar weights considered in literature, that ensures fast convergence and yields consistent estimates.  

It is worth noting here that there has been significant efforts in distributed parameter estimation via diffusion-based strategies~\cite{lopes2008diffusion,sayed2013diffusion}. The diffusion strategies aim at optimizing the aggregate MSE in a distributed manner and adapt combine-then-adapt (CTA) type of structure. Diffusion based estimation algorithms also considered deterministic parameter and has similar drawbacks as the contemporary related literature. 

Some of the important questions~\cite{rad2010distributed,bianchi2011convergence} in distributed algorithms include the algorithms' convergence pattern, consensus among the agents on their parameter estimates, distributed versus centralized performance comparison, and the rate of convergence. We discuss all these properties of our distributed estimators in the following sections. 

\section{Gaussian parameter estimation}
\label{sec-param-est}

\subsection{Problem Statement}
\label{subsec-problem}
Consider the problem of estimating a random vector parameter~$\btheta \in \mathbb{R}^{n}$ by a multi-agent network of~$m$ agents. The prior probability distribution of the parameter is Gaussian,~$\btheta \sim \mathcal{N}(\overline{\theta}, \Sigma_{\theta})$ where $\overline{\theta} \in \mathbb{R}^{n}$ and $\Sigma_{\theta} \in \mathbb{R}^{n\times n}$. 

Each agent~$i$ in the network observes only a few variables and makes low dimensional measurements~$\bz_{i,k} \in \mathbb{R}^{p_i}$, such that $p_i << n, \forall i=1,\hdots,m$ at iteration~$k$. The measurements are i.i.d and corrupted with noise, which we assume to be Gaussian. The observations of the agents in the cyber layer is represented by a linear and time-invariant model 
\begin{align}
	\label{eqn:obs}
	\bz_{i,k} = H_{i} \btheta + \bv_{i,k}, \qquad i=1,\hdots,m
\end{align}
where, $H_{i} \in \mathbb{R}^{p_i\times n}$ is the measurement matrix and~$\bv_{i,k} \in \mathbb{R}^{p_i}$ is the measurement noise. The Gaussian measurement noise, at each agent~$i$, has zero mean and covariance matrix $R_{i,k}$, i.e., $\bv_{i,k} \sim \mathcal{N}({\bm 0}_{p_i}, R_{i,k})$. The measurement noises~$\{\bv_{i,k}\}_{\forall i, k \geq 0}$ are uncorrelated random sequences.

The communication network helps the agents to share their measurements and current estimates with their neighbors. The communication network can be defined by a simple (no self-loops nor multiple edges) and \emph{directed} graph $\mathcal{G = (V,E)}$, where $\mathcal{V} = \{i: i=1,\hdots,m\}$ is the set of agents and  $\mathcal{E} = \{(i,j): \exists \text{ an edge }j \rightarrow i\}$ is the set of local communication channels among the agents. The communication graph is sparse and time-invariant. Note that the framework is based on a directed graph (one-way communications), that means the proposed approach is easily extendable to undirected graphs (two-way communications). The reverse extension is not always true. Let's denote the adjacency matrix of $\mathcal{G}$ by $\ad = [a_{ij}] \in \mathbb{R}^{m\times m}$, where, 
 \begin{align}
	     \label{eqn:net}
	     a_{ij} =  \left\{
	 \begin{array}{ll}
		 1,  & \mbox{if } \exists \text{ an edge }j \rightarrow i  \\
		 0, & \text{otherwise.}  \\
		 \end{array} \right. 
\end{align}
Readers are referred to \cite{chung1997spectral} for detailed information about graphs. The open and closed neighborhoods of each agent~$i$ are,
\begin{small}
	\begin{align}
		\label{eqn:open-nbhd}
		\Omega_i &= \{j|(i,j) \in  \mathcal{E}\} \\
		\label{eqn:closed-nbhd}
		\overline{\Omega}_i &=  \{i\} \cup \{j|(i,j) \in  \mathcal{E}\}.
	\end{align}
\end{small}
As considered in most distributed parameter estimation literature, the prior statistics of the parameter, $\overline{\theta}$ and $\Sigma_{\theta}$, the measurement models, $\{H_j, R_{j,k} : j \in \overline{\Omega}_i \}$, and the communication network model, $\mathcal{G}$ along with the adjacency matrix~$\ad$ are available to each agent~$i$ in the network. The prior statistics of the parameter, $\overline{\theta}$ and $\Sigma_{\theta}$, are used to set the initial condition of the distributed estimates. In problem scenarios where these statistics are not available, we provide an alternative initial conditions for the estimates, as discussed later. 

\subsection{Centralized Estimator}
\label{subsec-centralized}
Although not practical in the context of the problem considered in this paper, yet we lay down the centralized estimation algorithm to benchmark and compare the results on the distributed estimator. In a centralized scheme, all the agents in the cyber layer communicate their measurements to a central fusion center. The fusion center performs all needed computation tasks to obtain the estimate of the parameter~$\btheta$.

The aggregated measurements for a centralized centralized estimation scheme can be represented by,
\begin{align}
	\label{eqn:obs-cen}
	\bz_{k} &= H \btheta + \bv_{k}, \qquad \\
	\text{where, \quad} \bz_{k} &= \begin{bmatrix} \bz_{1,k} \\ \vdots \\  \bz_{m,k} \end{bmatrix}, \;\; H = \begin{bmatrix} H_1 \\ \vdots \\  H_m  \end{bmatrix}, \;\; \bv_{k} = \begin{bmatrix} \bv_{1,k} \\ \vdots \\  \bv_{m,k} \end{bmatrix}, \nonumber
\end{align}
i.e., $\bz_{k} \in \mathbb{R}^{p}$, $H \in \mathbb{R}^{p \times n}$, $\bv_{k} \in \mathbb{R}^{p}$, and~$p = \sum_{i=1}^{m}p_i$. The Gaussian measurement noise~$\bv_{k}$ still has zero mean and follows~$\bv_{k} \sim \mathcal{N}({\bm 0}_{p}, R_{k})$, where $R_k = \text{blockdiagonal}\{ R_1, \cdots, R_m \} \in \mathbb{R}^{p \times p}$.

Let $\xc_k \in \mathbb{R}^n$ denote the centralized estimate of~$\btheta$, $\pc_k \in \mathbb{R}^{n \times n}$ denote the estimation error covariance and $\kc_{k} \in \mathbb{R}^{n \times p}$ denote the gain matrix at iteration~$k$. At iteration~$k=0$, initialize the estimate and the error covariance\footnote{If the prior statistics of the parameter, $\overline{\theta}$ and $\Sigma_{\theta}$, are unknown, then the centralized estimator can be initialized with~$\xc_0 = {\bm 0}_n$ and $\pc_{0}=I_n$.} with~$\xc_0 = \overline{\theta}$ and $\pc_{0}=\Sigma_{\theta}$. The update equations of the centralized estimator for all~$k \geq 0$ are:
\begin{align}
	\label{eqn:gain-central}
	\kc_{k} &= \pc_{k}H^T \left( H\pc_{k}H^T + R_k  \right)^{-1} \\
	\label{eqn:est-update-central}
	\xc_{k+1} &= \xc_{k} + \kc_{k} \left( \bz_{k} - H \xc_{k} \right) \\
	\label{eqn:predcov-update-central}
	\pc_{k+1} &= \left( I_n - \kc_{k} H \right)\pc_{k}.
\end{align}
The equations~\eqref{eqn:gain-central}-\eqref{eqn:predcov-update-central} represent the centralized filter. The estimates~$\xc_{k}$ are essentially conditional means given the observations~$\bz_{k}$, i.e., 
\begin{align}
	\label{eqn:xc-mean}
	\xc_{k+1} &= \mathbb{E} \left[ \btheta \; | \; \{ \bz_{k} \}_{k = 0, \cdots, k}\right].
\end{align}

To analyze the convergence properties of the centralized estimator, we derive the expressions for the centralized estimator error~$\ec_k \in \mathbb{R}^n$ and innovation~$\yc_{k}\in \mathbb{R}^p$ sequences,
\begin{align}
	\label{eqn:error-central}
	\ec_{k} &= \btheta - \xc_{k} \\
	\label{eqn:inno-central}
	\yc_{k} &= \bz_{k} - H \xc_{k} \\
	&= H \btheta + \bv_{k} - H \xc_{k} \nonumber \\
	\label{eqn:inno-err-central}
	&= H \ec_{k} + \bv_{k}
\end{align}
The centralized error process follows the dynamics, utilizing~\eqref{eqn:est-update-central} and~\eqref{eqn:inno-err-central}, 
	\begin{align}
		\ec_{k+1} &= \btheta - \xc_{k+1} \nonumber \\
		&= \btheta - \xc_{k} - \kc_{k} \left( \bz_{k} - H \xc_{k} \right) \nonumber \\
		\label{eqn:err-inno-update-central}
		&= \ec_{k} - \kc_{k} \yc_{k} \\
		&= \ec_{k} - \kc_{k} \left( H \ec_{k} + \bv_{k} \right) \nonumber \\
		\label{eqn:err-dynamics-central}
		&= \left( I_n - \kc_{k} H \right)\ec_{k} - \kc_{k} \bv_{k}
	\end{align}
The convergence properties of the centralized estimator~\eqref{eqn:gain-central}-\eqref{eqn:predcov-update-central} is determined by the dynamics of the error process,~\eqref{eqn:err-dynamics-central}. If the error dynamics is asymptotically stable, then the error processes achieves asymptotic convergence that in turn guarantee the convergence of the estimation algorithm. The error dynamics is asymptotically stable if the spectral radius of the error's dynamics matrix is less than one, i.e.,~$\rho( I_n - \kc_{k} H ) < 1$. 
\begin{definition}[Centralized Observability]
	If the row rank of the centralized observability matrix~$H$ is equal to $n$, then the parameter is observable under the measurement model~\eqref{eqn:obs-cen}. This criteria is akin to the condition,
	\begin{align}
		\label{eqn:observability-central}
		\text{rank}\left( H^T H \right) = \text{rank}\left( \sum_{i=1}^{m} H_i^T H_i \right) = n.
	\end{align}
\end{definition}
Given that the observation model~\eqref{eqn:obs-cen} satisfies the centralized observability criteria~\eqref{eqn:observability-central}, it guarantees that there exists gain matrices~$\kc_{k}$ such that ~$\rho( I_n - \kc_{k} H ) < 1, \; \forall \: k \geq 0$.

The design of the optimal centralized gain matrices that ensures unbiasedness and consistency of the centralized estimates while ensuring fastest convergence, is obtained by applying Gauss-Markov theorem on
\begin{align}
	\xc_{k+1} = \xc_{k} + \kc_{k} \yc_{k}.
\end{align}
Note that both the centralized estimator error~$\ec_k$ and innovation~$\yc_{k}$ sequences are zero-mean, i.e., $\mathbb{E}[\ec_k] = 0$ and $\mathbb{E}[\yc_{k}] = 0$. Then the error covariance is~$\pc_k = \mathbb{E}[ \ec_k {\ec_k}^T]$. The centralized gain is constructed as:
	\begin{align}
		\label{eqn:central-gain-cov}
		\kc_{k} &= \Sigma_{\btheta \yc_k} \Sigma_{\yc_k}^{-1} \\
		\text{where, }& \nonumber \\
		\Sigma_{\btheta \yc_k} &= \mathbb{E}[ \left( \btheta - \overline{\theta} \right) \left( \yc_k - \mathbb{E}[\yc_{k}] \right)^T]  \nonumber \\
		&= \mathbb{E}[ \left( \btheta \!-\! \xc_{k} \!+\! \xc_{k} \!-\! \overline{\theta} \right) {\yc_k}^T] \nonumber \\
		&= \mathbb{E}[ \left( \btheta \!-\! \xc_{k} \right) {\yc_k}^T]  \nonumber \\
		&= \mathbb{E}[ \ec_k \left( H \ec_{k} \!+\! \bv_{k} \right)^T]  \nonumber \\
		\label{eqn:cross-cov-cen-gain}
		& = \! \pc_k H^T \\
		\Sigma_{\yc_k} &= \mathbb{E}[ \left( \yc_k - \mathbb{E}[\yc_{k}] \right) \left( {\yc_k} - \mathbb{E}[\yc_{k}] \right)^T ] \nonumber \\
		&= \mathbb{E}[ \left( H \ec_{k} + \bv_{k} \right) \left( H \ec_{k} + \bv_{k} \right)^T]  \nonumber \\
		\label{eqn:inno-cov-cen-gain}
		&= \! H \pc_k H^T + R_k.
	\end{align}
Replacing the covariance expressions from~\eqref{eqn:cross-cov-cen-gain} and~\eqref{eqn:inno-cov-cen-gain} into~\eqref{eqn:central-gain-cov} yields the optimal centralized estimator gains~\eqref{eqn:gain-central}. Leveraging the uncorrelated properties of the noise, innovation and error sequences it can be shown that $\mathbb{E}[ \left(\xc_{k} - \overline{\theta} \right) {\yc_k}^T] = 0$ and $\mathbb{E}[\ec_k \bv_{k}^T] = 0$ which are used in the derivation of the gain. 

Now we derive the recursive update of the centralized error covariance~$\pc_k$, starting with~\eqref{eqn:err-dynamics-central},
	\begin{align}
		\pc_{k+1} &= \mathbb{E}[ \ec_{k+1} {\ec_{k+1}}^T] \nonumber \\
		&= \mathbb{E}[ \left( \left( I_n \! - \!\kc_{k} H \right)\ec_{k} \!-\! \kc_{k} \bv_{k} \right) \left( \left( I_n \!-\! \kc_{k} H \right)\ec_{k} \!-\! \kc_{k} \bv_{k} \right)^T] \nonumber \\
		&= \left( I_n - \kc_{k} H \right) \pc_k \left( I_n - \kc_{k} H \right)^T + \kc_{k} R_k {\kc_{k}}^T \nonumber \\
		&= \pc_k - \kc_{k} H \pc_k - \pc_k H^T {\kc_{k}}^T + \kc_{k} H \pc_k H^T {\kc_{k}}^T + \kc_{k} R_k {\kc_{k}}^T \nonumber \\
		&= \pc_k \!-\! \kc_{k} H \pc_k \!-\! \pc_k H^T {\kc_{k}}^T \!\!+\! \kc_{k} \left( H \pc_k H^T \!\!+\! R_k \right) {\kc_{k}}^T \nonumber \\
		&= \pc_k - \kc_{k} H \pc_k - \pc_k H^T {\kc_{k}}^T + \pc_k H^T{\kc_{k}}^T \nonumber \\
		\label{eqn:error-cov-recursion-central}
		&= \left( I_n - \kc_{k} H \right) \pc_k,
	\end{align}
which is the update equation~\eqref{eqn:predcov-update-central}. This completes the derivation of all the processes of the centralized estimator. 

Since the centralized framework satisfies the centralized observability criteria~\eqref{eqn:observability-central}, the gain matrices~$\kc_k$ in~\eqref{eqn:gain-central} are such that~$\rho( I_n - \kc_{k} H ) < 1 \; \forall \: k \geq 0$. In the limit, the centralized error covariance then decays down to a zero matrix,
\begin{align}
	\lim_{k \rightarrow \infty} \pc_{k} &= \lim_{k \rightarrow \infty} \left( I_n - \kc_{k-1} H \right) \pc_{k-1} \nonumber \\
	&= \displaystyle \left( \lim_{k \rightarrow \infty} \Pi_{k = 0}^{k-1} \left( I_n - \kc_{k} H \right) \right) \pc_{0} = {\bm 0}_{n \times n}.
\end{align}
Thus, as~$k \rightarrow \infty$ the centralized estimate~$\xc_k$ converges to the true parameter vector~$\btheta$, i.e.,~$\mathbb{P} \begin{bmatrix} \displaystyle \lim_{k \rightarrow \infty} \xc_k = \btheta \end{bmatrix} = 1$. Hence the centralized estimator is a consistent estimator.

\section{Distributed parameter estimation}
\label{sec-distributed-estimation}
%
In this paper, a distributed approach is proposed that tracks the centralized approach at each agent. For each agent to be able to successfully track the centralized estimator, a key condition is the distributed observability that needs to be satisfied. 

\subsection{Distributed observability for parameter estimation}
\label{subsec-obs}
This paper introduces a new notion of distributed observability for parameter estimation by extending the distributed observability definition provided in~\cite{das2022observability} for distributed state estimation. The distributed observability at each agent is a measure of how well the parameter can be inferred from knowledge of its local measurements and interactions among agents in the network. 

Consider the observation-communication model represented by equations~\eqref{eqn:obs} and ~\eqref{eqn:net}. The \emph{connectivity matrix}~$\tilde{\ad}$ of the network is defined (in~\cite{das2022observability}), in terms of adjacency matrix $\ad$,  as
\begin{align}
	\label{eqn:connectivity-matrix}
	\tilde{\ad} = I_m + \ad + \ad^2 + \hdots + \ad^{m-1}.
\end{align}
As described in~\cite{das2022observability}, the element~$[\ad^q]_{i,j}$ of the matrix,~$\ad^q \; \forall q \in \mathcal{Z}^+$, gives the number of directed walks of length~$q$ from agent~$j$ to agent~$i$. Then, the connectivity matrix is a non-negative matrix\footnote{If there exists $i,j$ such that $\tilde{a}_{i,j}=0$, then there doesn't exist any path from $j$ to $i$ and would imply that the graph is not connected. The agent communication network, i.e., the directed graph, is connected if the connectivity matrix, defined in~\eqref{eqn:connectivity-matrix}, is a positive matrix, i.e., $\tilde{\ad}>0$. For a \emph{fully} connected network, $I_m + \ad > 0$.},~$\tilde{\ad} \geq 0$, and its elements~$[\tilde{\ad}]_{i,j}=\tilde{a}_{i,j}$ denote the total number of walks (of any length $<m$) from node $j$ to node $i$.

Let the $\tilde{\ad}_i$ denote the~$i^{\text{th}}$ row of the matrix~$\tilde{\ad}$ and the symbol~$\bullet$ denote the face-splitting product of matrices (transposed Khatri–Rao product).
\begin{definition}[Distributed Observability - Parameter estimation]
	If the row rank of the distributed observability matrix~$\mathcal{O}_i$, defined as,
	\begin{align}
		\label{eqn:dist-obs}
		\mathcal{O}_i = \tilde{\ad}_i \bullet H = 
		\begin{bmatrix} \tilde{a}_{i,1} \\ \tilde{a}_{i,2} \\ \vdots \\ \tilde{a}_{i,m} \end{bmatrix} \bullet
		\begin{bmatrix} H_1 \\ H_2 \\ \vdots \\ H_m \end{bmatrix}
		= \begin{bmatrix} \tilde{a}_{i,1} H_1 \\ \tilde{a}_{i,2} H_2 \\ \vdots \\ \tilde{a}_{i,m} H_m \end{bmatrix}
	\end{align}
	is equal to $n$, then the parameter is distributedly observable at agent~$i$. This condition is equivalent to,
	\begin{align}
		\text{rank}( \mathcal{O}_i^T \mathcal{O}_i ) = \text{rank}\left( \sum_{j=1}^{m} \tilde{a}_{i,j}^2 H_j^T H_j \right) = n
	\end{align}
\end{definition}

Most of the papers in distributed parameter estimation assumes that both the centralized observability~\eqref{eqn:observability-central} condition holds and that the graph is connected. In contrast, this distributed observability for parameter estimation is a weaker assumption and does not require the graph to be connected. The distributed parameter estimation algorithm proposed in this paper needs the underlying framework to only satisfy the distributed observability criteria and is thus a more broadly applicable approach.

\subsection{Distributed Parameter Estimation Algorithm}
\label{subsec-algo}

At time~$k$ and agent~$i$, let $\bx_{i,k} \in \mathbb{R}^n$ denote the distributed estimate of~$\btheta$, $\bp_{i,k} \in \mathbb{R}^{n \times n}$ denote the estimation error covariance, $\bp_{ij,k} \in \mathbb{R}^{n \times n}$ denote the cross-covariance between estimation errors of agent~$i$ \& agent~$j$, and $\bk_{i,k} \in \mathbb{R}^{ n \times \left( \sum_{j \in \cn_i}p_j + n|\on_i| \right)}$ denote the gain matrix. At iteration~$k=0$, initialize the estimate and the error covariance\footnote{If the prior statistics of the parameter, $\overline{\theta}$ and $\Sigma_{\theta}$, are unknown, then the centralized estimator can be initialized with~$\bx_{i,0} = {\bm 0}_n$, $\bp_{i,0}=I_n$, and $\bp_{ij,0}=I_n \; \forall \; i=1,\cdots,m$ and $j \in \on_i$. Note that we denote $\bp_{ii,k}$ by $\bp_{i,k}$ for ease of notation.} at each agent~$i$ with~$\bx_{i,0} = \overline{\theta}$, $\bp_{i,0}=\Sigma_{\theta}$, $\bp_{ij,0}=\Sigma_{\theta} \; \forall j \in \on_i$, and $\bp_{ij,k}=  {\bm 0}_{n\times n} \; \forall k, j \not\in \cn_i$. The update equations of the distributed estimation algorithm at each agent~$i$ for all~$k \geq 0$ are:
\begin{align}
	\label{eqn:gain-update}
	\bk_{i,k} &= \Sigma_{\btheta, y_i} \Sigma^{-1}_{y_i} \\
	\label{eqn:x-update}
	\bx_{i,k\!+\!1} &\! = \! \bx_{i,k} \!+\!\!\! \sum_{j \in \Omega_i}\!\! B_{ij,k} \!\! \left( \bx_{j,k} \!-\! \bx_{i,k}\right) \!+ \!\!\!\sum_{j \in \overline{\Omega}_i} \!\! M_{ij,k} \!\! \left( \bz_{j,k} \!-\! H_j \bx_{i,k} \right) \\
	\label{eqn:p-update}
	\bp_{i,k+1} &= \bp_{i,k} - \bk_{i,k} \Sigma_{\btheta, y_i}^T \\
	\label{eqn:cp-update}
	\bp_{ij,k\!+\!1} &\!=\! \bp_{ij,k} \!-\! \bk_{i,k} \Sigma_{\btheta, y_i}^T \!\!-\! \Sigma_{\btheta, y_j}\bk_{j,k}^T \!+\! \bk_{i,k} \Sigma_{y_{ij}}\! \bk_{j,k}^T, \quad \forall j \in \on_i
\end{align}
where, $j \in \on_i$, the covariance matrices~$\Sigma_{\btheta, y_i}, \Sigma_{y_i}$ and~$\Sigma_{y_{ij}}$ are derived as part of the optimal gain design in subsection~\ref{subsec:gain}. The parameter estimate update step~\eqref{eqn:x-update} is of \emph{consensus+innovations} type\footnote{The distributed estimator is inspired by the pseudo-innovations 
	approaches that are summarized in~\cite{das2013ICASSP, das2013EUSIPCO}}, where~$B_{ij,k} \in \mathbb{R}^{n \times n}$ are the local consensus weight matrices and~$M_{ij,k} \in \mathbb{R}^{n \times p_j}$ are the local innovation weight matrices. 

It is to be noted that not only the observations~$\{\bz_{j,k}\}_{j \in \cn_i}$ but also the parameter estimates~$\{\bx_{j,k} \}_{j \in \on_i}$ from the neighbors can be considered as additional information at each agent at each iteration. Based on this characterization, the notions of consensus and innovations are fused together into innovations~$\by_{i,k}\in \mathbb{R}^{\left( \sum_{j \in \cn_i}p_j + n|\on_i| \right)}$ at each agent~$i$ as represented in the following equation,
	\begin{align}
		\label{eqn:x-inno}
		\bx_{i,k+1} &= \bx_{i,k} + \bk_{i,k} \underbrace{\begin{bmatrix} 
				\bz_{j_1,k} - H_{j_1} \bx_{i,k} \\ \cdot\cdot \\ \bz_{i,k} - H_i \bx_{i,k} \\ \cdot \cdot \\ \bz_{j_{|\cn_i|},k} - H_{j_{|\cn_i|}} \bx_{i,k} \\
				\bx_{j_1,k} - \bx_{i,k} \\ \vdots \vdots \\ \bx_{j_{|\on_i|},k} - \bx_{i,k}
		\end{bmatrix}}_{\by_{i,k}},  \quad \text{where,}\\
		\label{eqn:gain2weights}
		\bk_{i,k} \! =& \! \begin{bmatrix} \! M_{ij_1,k}, \!\cdot \cdot, \! M_{ii,k}, \! \cdot \cdot, \! M_{ij_{|\cn_i|},k}, \! B_{ij_1,k}, \cdots, B_{ij_{|\on_i|},k} \! \end{bmatrix} \!, 
	\end{align}
$\{j_1, \cdot \cdot, i, \cdot \cdot, j_{|\cn_i|} \} = \cn_i$ and  $\{j_1, \cdots, j_{|\on_i|}\} = \on_i$.
The local consensus~$B_{ij,k}$ and local innovation~$M_{ij,k}$ weight matrices in the parameter estimate update step~\eqref{eqn:x-update} are obtained from the distributed gain matrix~$\bk_{i,k}$ by utilizing the expression in~\eqref{eqn:gain2weights}. The equations~\eqref{eqn:gain-update}-\eqref{eqn:p-update} along with~~\eqref{eqn:gain2weights} represents the proposed distributed parameter estimation algorithm that fuses the concepts of consensus and innovations by treating the consensus on the state estimates as innovations along with the local innovations of the agent and its neighbors. In the next section, we investigate the convergence properties of this algorithm and provide a design of the optimal gain matrix that ensures fast convergence. 

%
%

\section{Convergence Analysis}
\label{sec:convergence}
Similar to the centralized estimator, the distributed estimates~$\bx_{i,k+1}$ are conditional means of the parameter~$\btheta$ given the observations~$\{\bz_{j,k}\}_{j \in \cn_i}$ and the neighbors' parameter estimates~$\{\bx_{j,k} \}_{j \in \on_i}$, i.e., 
%
	\begin{align}
		\label{eqn:x-mean}
		\bx_{i,k+1} \! = \!  \mathbb{E} \left[ \btheta \; | \; \{ \bz_{j,k} \}_{j \in \cn_i}, \{ \bx_{j,k} \}_{j \in \on_i}, k = 0, \cdots, k \right].
	\end{align}
%
Using this relation, it can be shown that the innovation sequences~$\{\by_{i,k}\}_{\forall i, k\geq0}$ are Gaussian random vectors, uncorrelated and are with zero mean, $\mathbb{E}[\by_{i,k}]=0$, $\forall i, k\geq 0$. Building from these expressions, we analyze the error processes of the distributed estimation algorithm.

\subsection{Error Analysis}
\label{subsec:error}
The distributed estimation error terms,~$\ed_{i,k} \in \mathbb{R}^n$ at each agent~$i$, are defined as,
\begin{align}
	\label{eqn:ed}
	\ed_{i,k} = \btheta - \bx_{i,k}.
\end{align}
The error processes~$\ed_{i,k}$ are unbiased, i.e, they are zero mean at all agents and for all time indices, $\mathbb{E}[\ed_{i,k}]=0$, $\forall i, k\geq 0$. The error processes follow: $\ed_{i,k} \sim \mathcal{N} ({\bm 0}_n, \bp_{i,k})$. We now derive the innovations in terms of the error process and vice versa. First, using equation~\eqref{eqn:x-inno} and using~\eqref{eqn:obs}, the innovations are expanded as:
%
	\begin{align}
		\by_{i,k} \!\!&=\!\! \begin{bmatrix} 
			H_{j_1} \btheta + \bv_{j_1,k} - H_{j_1} \bx_{i,k} \\ \cdot\cdot \\  H_{i} \btheta + \bv_{i,k} - H_i \bx_{i,k} \\ \cdot \cdot \\  H_{j_{|\cn_i|}} \btheta \!+\! \bv_{j_{|\cn_i|},k} \!-\! H_{j_{|\cn_i|}} \bx_{i,k} \\
			\bx_{j_1,k} - \btheta + \btheta - \bx_{i,k} \\ \vdots \vdots \\ \bx_{j_{|\on_i|},k} - \btheta + \btheta - \bx_{i,k}
		\end{bmatrix}
		\!\! = \!\!  \underbrace{ \begin{bmatrix} 
				H_{j_1} \\ \cdot\cdot \\  H_{i} \\ \cdot \cdot \\  H_{j_{|\cn_i|}} \\ I_n \\ \vdots \vdots \\ I_n \end{bmatrix} }_{\tilde{H}_i} \! \ed_{i,k} \!\!+\!\! 
		\underbrace{ \begin{bmatrix} 
				\bv_{j_1,k}\\ \cdot\cdot \\  \bv_{i,k} \\ \cdot \cdot \\  \bv_{j_{|\cn_i|},k} \\
				- \ed_{j_1,k} \\ \vdots \vdots \\ -  \ed_{j_{|\on_i|},k}
		\end{bmatrix} }_{\delta_{i,k}} \nonumber \\
		\label{eqn:inno-dynamics}
		\by_{i,k} \!\! &= \tilde{H}_i \ed_{i,k} + \delta_{i,k} \qquad \forall i, k\geq 0,
	\end{align}
%
where,~$\tilde{H}_i \in \mathbb{R}^{ \left( \sum_{j \in \cn_i}p_j + n|\on_i| \right) \times n}$ are the local innovation matrices and the~$\delta_{i,k} \in \mathbb{R}^{\sum_{j \in \cn_i}p_j + n|\on_i|}$ are the local innovation noises at each agent~$i$. The dynamics of the local innovation processes are represented in compact notation by~\eqref{eqn:inno-dynamics}. 
The local innovation noises~$\delta_{i,k}$ are Gaussian random vectors with zero mean and let the variance be denoted by~$\Delta_{i,k}$, i.e., $\delta_{i,k} \sim \mathcal{N}({\bm 0}, \Delta_{i,k})$.

The distributed error process~$\ed_{i,k}$ follows the dynamics, utilizing~\eqref{eqn:x-inno} and~\eqref{eqn:inno-dynamics},
%
	\begin{align}
		\label{eqn:ed-inno-dyn}
		\! \! \!  \ed_{i,k+1} &= \btheta \! - \! \bx_{i,k+1} \nonumber\\
		 & =  \btheta \! - \! \left( \bx_{i,k} \! + \! \bk_{i,k} \by_{i,k} \right) \nonumber\\
		& =  \ed_{i,k} \! - \! \bk_{i,k} \by_{i,k} \\
		&= \ed_{i,k} - \bk_{i,k} \left( \tilde{H}_i \ed_{i,k} + \delta_{i,k} \right)  \nonumber \\
		\label{eqn:ed-dyn}
		&= \left( I_n - \bk_{i,k} \tilde{H}_i \right)\ed_{i,k} - \bk_{i,k}\delta_{i,k}, 
	\end{align}
%
and, the evolution of the distributed error covariance~$\bp_{i,k}$ takes the form, using~\eqref{eqn:ed-inno-dyn},
%
	\begin{align}
		\bp_{i,k+1} &\!= \!\mathbb{E}[\ed_{i,k+1} \ed_{i,k+1}^T]  \nonumber\\
		& = \mathbb{E} \left[ \left( \ed_{i,k} \!-\! \bk_{i,k} \by_{i,k} \right) \left( \ed_{i,k} \!-\! \bk_{i,k} \by_{i,k} \right)^T \right] \nonumber\\
		&= \bp_{i,k} + \bk_{i,k} \Sigma_{y_i} \bk_{i,k}^T - \Sigma_{\btheta, y_i} \bk_{i,k}^T - \bk_{i,k} \Sigma_{\btheta, y_i}^T \nonumber\\
		&= \bp_{i,k} \!+\! \bk_{i,k} \Sigma_{y_i} \left( \Sigma_{\btheta, y_i} \Sigma^{-1}_{y_i} \right)^T \!\! - \! \Sigma_{\btheta, y_i} \bk_{i,k}^T \!-\! \bk_{i,k} \Sigma_{\btheta, y_i}^T \nonumber\\
		\label{eqn:p-update-deriv}
		&= \bp_{i,k} - \bk_{i,k} \Sigma_{\btheta, y_i}^T.
	\end{align}
%
The term~$\mathbb{E}[ \ed_{i,k} \by_{i,k}^T] = \Sigma_{\btheta, y_i}$ as shown later in~\eqref{eqn:equi-cross-cov} and by substituting~$\bk_{i,k}$ with~$\Sigma_{\btheta, y_i} \Sigma^{-1}_{y_i}$ we obtain equation~\eqref{eqn:p-update-deriv}.  Similarly, the distributed error cross covariance~$\bp_{ij,k}$ is derived as,
%
\begin{align}
	\bp_{ij,k+1} &= \mathbb{E}[\ed_{i,k+1} \ed_{j,k+1}^T] \nonumber\\
	&= \mathbb{E} \left[ \left( \ed_{i,k} \!-\! \bk_{i,k} \by_{i,k} \right) \left( \ed_{j,k} \!-\! \bk_{j,k} \by_{j,k} \right)^T \right] \nonumber\\
	\label{eqn:p-cross-update-deriv}
	&= \bp_{ij,k} - \bk_{i,k} \Sigma_{\btheta, y_i}^T - \Sigma_{\btheta, y_j} \bk_{j,k}^T + \bk_{i,k} \Sigma_{y_{ij}} \bk_{j,k}^T .
\end{align}
In the above derivation,~$\mathbb{E}[ \ed_{j,k} \by_{i,k}^T]$ also reduces to $\Sigma_{\btheta, y_i}$ and the relation holds true when~$i$ and~$j$ are interchanged. The designs of the gain matrices and the covariances are provided in the next subsection \eqref{eqn:equi-cross-cov}-\eqref{eqn:delta-cov-cross}. Equations~\eqref{eqn:p-update-deriv} and \eqref{eqn:p-cross-update-deriv} are the update steps for the distributed error covariances of the proposed distributed parameter estimation algorithm.

\subsection{Optimal Gain Design}
\label{subsec:gain}
Following along the same lines of centralized optimal gain matrix design, we state that the optimal distributed gain matrices that ensures unbiasedness and consistency of the distributed estimates while ensuring fast convergence, are obtained by following the first principles of Gauss-Markov theorem.
Since the zero-mean innovation sequences~$\{\by_{i,k}\}_{\forall i, k\geq0}$ are Gaussian and uncorrelated, they are independent random vectors. By applying Gauss-Markov theorem on equation~\eqref{eqn:x-inno}, the optimal gain matrices for the distributed parameter estimates are constructed as:
\begin{align}
	\bk_{i,k} &= \Sigma_{\btheta, y_i} \Sigma^{-1}_{y_i}, \qquad \text{where,} \nonumber \\
	\Sigma_{\btheta, y_i} &= \mathbb{E}[ \left( \btheta - \overline{\theta} \right)\by_{i,k}^T] \nonumber \\
	\label{eqn:equi-cross-cov}
	& = \mathbb{E}[ \left( \btheta - \bx_{i,k} + \bx_{i,k} - \overline{\theta} \right) \by_{i,k}^T] = \mathbb{E}[ \ed_{i,k}\by_{i,k}^T]  \\
	\label{eqn:cross-inno-x}
	&=  \mathbb{E}[ \ed_{i,k}( \tilde{H}_i \ed_{i,k} + \delta_{i,k} )^T] 
	= \bp_{i,k}\tilde{H}_i^T + \Sigma_{\epsilon_i, \delta_i}\\
	\text{and,}& \nonumber \\ 
	\Sigma_{y_i} &= \mathbb{E} [\by_{i,k} \by_{i,k}^T] 
	= \mathbb{E} [ (\tilde{H}_i \ed_{i,k} + \delta_{i,k}) (\tilde{H}_i \ed_{i,k} + \delta_{i,k})^T ] \nonumber \\
	&= \tilde{H}_i \bp_{i,k}\tilde{H}_i^T + \Delta_{i,k} + \tilde{H}_i \Sigma_{\epsilon_i, \delta_i} + \Sigma_{\epsilon_i, \delta_i}^T \tilde{H}_i^T.
\end{align}

The fact that~$\mathbb{E}\left[ \left( \bx_{i,k} - \overline{\theta} \right) \by_{i,k}^T \right] =0$ was utilized to obtain~\eqref{eqn:equi-cross-cov}, which can be shown by using techniques similar to the ones presented in~\cite{das2015TSP,das2017TSP}. The covariance of the innovation noise process~$\Delta_{i,k}$ and its cross covariance with the distributed error process~$\Sigma_{\epsilon_i, \delta_i}$ take the forms:
	\begin{align}
		\!\! \Sigma_{\epsilon_i, \delta_i} &\!\!= \mathbb{E}[ \ed_{i,k}\delta_{i,k}^T] \nonumber \\
		\label{eqn:sig-eps-delta}
		&\!\! = \begin{bmatrix}
			{\bm 0}_{n,p_{\!_{j_1}}} \!\!\!\!\!\!\!&\!\! \cdot\cdot \!\!\!&\!\!  {\bm 0}_{n,p_{i}} \!\!\!&\!\!\!\! \cdot \cdot \!\!\!&\!\!\!  {\bm 0}_{n,p_{j_{\!_{|\cn_i|}}}} \! \vdots -\! \bp_{ij_{\!_1},k} \cdots -\! \bp_{ij_{\!_{|\on_i|}},k}
		\end{bmatrix} \\
		\Delta_{i,k} &= \mathbb{E}[ \delta_{i,k} \delta_{i,k}^T] \nonumber \\
		\label{eqn:delta-cov}
		& = \text{blkdiag}\begin{Bmatrix} \text{blkdiag}\{ R_{j,k} \}_{j \in \cn_i},  \left[ \left[ \bp_{jl,k} \right]_{j,l \in \on_i} \right] \end{Bmatrix}
	\end{align}
where, blkdiag means a block-diagonal matrix and we leverage the relations that~$\mathbb{E}[\ed_{i,k}\bv_{j,k}^T] =0 \; \forall\; j \in \cn_i$, to derive the above two covariance quantities. 
The only term left is the innovation cross-covariance,~$\Sigma_{y_{ij}}$, which can be derived similarly using the steps described above as,
%
\begin{align}
\Sigma_{y_{ij}} &= \mathbb{E} [\by_{i,k} \by_{j,k}^T] \nonumber \\
&= \mathbb{E} [ (\tilde{H}_i \ed_{i,k} + \delta_{i,k}) (\tilde{H}_j \ed_{j,k} + \delta_{j,k})^T ] \nonumber \\
&= \tilde{H}_i \bp_{ij,k}\tilde{H}_i^T + \Delta_{ij,k} + \tilde{H}_i \Sigma_{\epsilon_i, \delta_j} + \Sigma_{\epsilon_j, \delta_i}^T \tilde{H}_j^T, \\
\text{where, } \qquad \qquad &\nonumber \\
	\!\! \Sigma_{\epsilon_i, \delta_j} &\!\!= \mathbb{E}[ \ed_{i,k}\delta_{j,k}^T] \nonumber \\
\label{eqn:sig-eps-delta-cross}
&\!\! = \begin{bmatrix}
	{\bm 0}_{n,p_{\!_{l_1}}} \!\!\!\!\!\!\!&\!\! \cdot\cdot \!\!\!&\!\!  {\bm 0}_{n,p_{j}} \!\!\!&\!\!\!\! \cdot \cdot \!\!\!&\!\!\!  {\bm 0}_{n,p_{l_{\!_{|\cn_j|}}}} \! \vdots -\! \bp_{il_{\!_1},k} \cdots -\! \bp_{il_{\!_{|\on_j|}},k}
\end{bmatrix} \\
\Delta_{ij,k} &= \mathbb{E}[ \delta_{i,k} \delta_{j,k}^T] \nonumber \\
\label{eqn:delta-cov-cross}
& = \text{blkdiag}\begin{Bmatrix} \text{blkdiag}\{ R_{ql,k} \}_{q \in \cn_i, l \in \cn_j},  \left[ \left[ \bp_{ql,k} \right]_{q \in \on_i, l \in \on_j} \right] \end{Bmatrix}.
\end{align}
%
Since the observation noises~$\bv_{i,k}$ are uncorrelated,~$R_{ql,k} = R_q$ if $q=l$ otherwise~$R_{ql,k} = 0_{p_q, p_l}$.
Note that the innovation noise covariance and cross covariance are fully expressed in terms of the observation noise covariances and the estimation error cross covariances. With the two expressions~$\Sigma_{\epsilon_i, \delta_i}$ and~$\Delta_{i,k}$ in~\eqref{eqn:sig-eps-delta}-\eqref{eqn:delta-cov}, the optimal gain matrices at each agent for the distributed estimator expands into:
%
	\begin{align}
		\label{eqn:gain-full}
		\bk_{i,k} = \left( \bp_{i,k}\tilde{H}_i^T + \Sigma_{\epsilon_i, \delta_i} \right) \Big( \tilde{H}_i \bp_{i,k}\tilde{H}_i^T + \Delta_{i,k} + \tilde{H}_i \Sigma_{\epsilon_i, \delta_i}  + \Sigma_{\epsilon_i, \delta_i}^T \tilde{H}_i^T \Big)^{-1}
	\end{align}
%
Note that the gain matrices will be very sparse at each agent. To alleviate challenges in tracking of the complete network error covariances,~\cite{sebastian2021all} presents a certifiable optimal distributed filter that performs optimal fusion of estimates under unknown correlations by a particular tight Semidefinite Programming (SDP) relaxation. Further, given that the matrices does not depend on the measurements, they all can be pre-computed and stored at each agent.

\subsection{Asymptotic properties}
\label{subsec:asymp}

The convergence properties of distributed parameter estimation algorithm~\eqref{eqn:gain-update}-\eqref{eqn:p-update} can be investigated through the lens of the behavior of the error dynamics~\eqref{eqn:ed-dyn}. If the error dynamics are asymptotically stable, i.e., the spectral radius of the error's dynamics matrix is less than one, ~$\rho\left( I_n - \bk_{i,k} \tilde{H}_i \right) < 1, \; \forall \; i,$ then the error processes have asymptotically decaying error covariances that in turn guarantee the convergence of the distributed algorithm.

Given that the \emph{observation-network} model~\eqref{eqn:obs}-\eqref{eqn:net} satisfies the Distributed Observability criteria~\eqref{eqn:dist-obs} for parameter estimation, it guarantees that there exists gain matrices~$\bk_{i,k}$ at each agent~$i$ such that
\begin{align}
	\label{eqn:rho_dist}
	\rho\left( I_n - \bk_{i,k} \tilde{H}_i \right) < 1 \; \forall \; i.
\end{align}
Note that there may exist multiple realizations of the distributed gain matrices~$\bk_{i,k} \in \mathbb{R}^{ n \times \left( \sum_{j \in \cn_i}p_j + n|\on_i| \right)}$ satisfying~\eqref{eqn:rho_dist} and for all such realizations the convergence of the distributed algorithm is guaranteed. The optimal design of the distributed gain matrices, provided in this paper \eqref{eqn:gain-update} \eqref{eqn:gain2weights} and \eqref{eqn:gain-full}, not only guarantees convergence by satisfying~\eqref{eqn:rho_dist} but also ensures fastest convergence.

From equations~\eqref{eqn:p-update-deriv} and~\eqref{eqn:cross-inno-x}, we study the asymptotic behavior of the distributed estimation error covariance,  
	\begin{align}
		\lim_{k \rightarrow \infty} \bp_{i,k+1} &= \lim_{k \rightarrow \infty} \bp_{i,k} - \bk_{i,k} \left( \bp_{i,k}\tilde{H}_i^T + \Sigma_{\epsilon_i, \delta_i} \right)^T \nonumber \\
		&= \lim_{k \rightarrow \infty}  \left( I_n - \bk_{i,k} \left( \tilde{H}_i + \Sigma_{\epsilon_i, \delta_i}^T \bp_{i,k}^{-1} \right) \right) \bp_{i,k} \nonumber \\
		\label{eqn:cov-limit}
		&= \left( \lim_{k \rightarrow \infty}  \Pi_{k = 0}^{k} \left( I_n - \bk_{i,k} \left( \tilde{H}_i + \Sigma_{\epsilon_i, \delta_i}^T \bp_{i,k}^{-1} \right) \right) \right) \bp_{i,0}.
	\end{align}
We can prove (by contradiction) that the design of the gain matrices~$\bk_{i,k}$ are such that $\rho\left( I_n - \bk_{i,k} \left( \tilde{H}_i + \Sigma_{\epsilon_i, \delta_i}^T \bp_{i,k}^{-1} \right) \right) < 1 \; \forall \; i$ at all time indices $k \geq 0$. Let's assume that the spectral radius is not less than unity for some~$i = \iota$ and~$k=\kappa$, then we replace~$\bk_{\iota,\kappa}$ with 
\begin{align}
	\tilde{\bk}_{\iota,\kappa} = \begin{bmatrix} \tilde{M}_{\iota j_1,\kappa}, \cdot \cdot, \tilde{M}_{\iota \iota,\kappa}, \cdot \cdot, \tilde{M}_{\iota j_{|\cn_{\iota}|},\kappa}, \tilde{B}_{\iota j_1,\kappa}, \cdots, \tilde{B}_{\iota j_{|\on_{\iota}|},\kappa} \end{bmatrix} \nonumber \\
	\text{where},\;\;  \tilde{M}_{\iota j,\kappa} = M_{\iota j,\kappa} \; \forall \; j \in \cn_{\iota}, \; \text{and,} \; \tilde{B}_{\iota j,\kappa} = B_{ \iota j,\kappa} \left( I_n - \bp_{ \iota j,\kappa} \bp_{ \iota,\kappa}^{-1} \right), \forall \; j \in \on_{\iota} \nonumber. 
\end{align}
Once the term~$\tilde{H}_i + \Sigma_{\epsilon_i, \delta_i}^T \bp_{i,k}^{-1}$ in equation~\eqref{eqn:cov-limit} is expanded using the definition of $\Sigma_{\epsilon_i, \delta_i}$ from~\eqref{eqn:sig-eps-delta}, the spectral radius $\rho\left( I_n - \tilde{\bk}_{\iota,\kappa} \left( \tilde{H}_{\iota} + \Sigma_{\epsilon_{\iota}, \delta_{\iota}}^T \bp_{\iota,\kappa}^{-1} \right) \right) $ becomes less than unity, given that $\rho\left( I_n - \bk_{\iota,\kappa} \tilde{H}_{\iota} \right) < 1$ is true because of the distributed observability assumption. 

Since the spectral radius of the product terms in the right hand side of~\eqref{eqn:cov-limit} are all less than unity, then in the limit, as $k \rightarrow \infty$, the distributed estimation error covariance~$\bp_{i,k}$ reduces to zero, i.e., $\lim_{k \rightarrow \infty} \bp_{i,k+1} = 0, \; \forall \; i$. 
Hence, similar to its centralized counterpart, the proposed distributed estimator is also a consistent estimator, i.e., the distributed  estimates~$\bx_{i,k}$ converge to the true parameter vector~$\btheta$, i.e.,~$\mathbb{P} \begin{bmatrix} \displaystyle \lim_{k \rightarrow \infty} \bx_{i,k} = \btheta \end{bmatrix} = 1$ at all the agents~$i$ in the network.



\section{Conclusions}
This paper considers the distributed parameter estimation problem where the parameter is stochastic and Gaussian distributed. The primary contributions are: (a) introducing a new distributed parameter estimation algorithm that outperforms state-of-the-art approaches by incorporating consensus on neighbors' estimates as innovations; (b) defining a new distributed parameter observability criteria that is a weaker assumption compared to the assumptions in the literature; and, (c) most importantly, designing the gain matrices, instead of scalar weights, for the distributed estimator such that the algorithm is optimal, i.e., it yields consistent estimates at all agents and achieves fast convergence. The framework and the methodology proposed in this paper will serve as a fundamental backbone for several downstream research problems in distributed (stochastic) parameter estimation, especially for optimal sensor placement, adaptation to node or communication failures.

\bibliographystyle{unsrt}
\bibliography{references}

\begin{thebibliography}{10}

\bibitem{chong2017forty}
Chee-Yee Chong.
\newblock Forty years of distributed estimation: A review of noteworthy
  developments.
\newblock In {\em 2017 Sensor Data Fusion: Trends, Solutions, Applications
  (SDF)}, pages 1--10. IEEE, 2017.

\bibitem{ding2019survey}
Derui Ding, Qing-Long Han, Zidong Wang, and Xiaohua Ge.
\newblock A survey on model-based distributed control and filtering for
  industrial cyber-physical systems.
\newblock {\em IEEE Transactions on Industrial Informatics}, 15(5):2483--2499,
  2019.

\bibitem{ge2019distributed}
Xiaohua Ge, Qing-Long Han, Xian-Ming Zhang, Lei Ding, and Fuwen Yang.
\newblock Distributed event-triggered estimation over sensor networks: A
  survey.
\newblock {\em IEEE transactions on cybernetics}, 50(3):1306--1320, 2019.

\bibitem{kar2012distributed}
S.~Kar, José M.~F. Moura, and K.~Ramanan.
\newblock Distributed parameter estimation in sensor networks: Nonlinear
  observation models and imperfect communication.
\newblock {\em IEEE Transactions on Information Theory}, 58(6):3575--3605,
  2012.

\bibitem{karmoura-spm2013}
S.~Kar and Jos{\'e} M.~F. Moura.
\newblock Consensus+innovations distributed inference over networks:
  cooperation and sensing in networked systems.
\newblock {\em IEEE Signal Processing Magazine}, 30(3):99--109, May 2013.

\bibitem{kar2011convergence}
S.~Kar and José M.~F. Moura.
\newblock Convergence rate analysis of distributed gossip (linear parameter)
  estimation: Fundamental limits and tradeoffs.
\newblock {\em IEEE Journal of Selected Topics in Signal Processing},
  5(4):674--690, 2011.

\bibitem{konevcny2016federated}
Jakub Kone{\v{c}}n{\`y}, H~Brendan McMahan, Felix~X Yu, Peter Richt{\'a}rik,
  Ananda~Theertha Suresh, and Dave Bacon.
\newblock Federated learning: Strategies for improving communication
  efficiency.
\newblock {\em arXiv preprint arXiv:1610.05492}, 2016.

\bibitem{ribeiro2004distributed}
Alejandro Ribeiro and Georgios~B Giannakis.
\newblock Distributed estimation in gaussian noise for bandwidth-constrained
  wireless sensor networks.
\newblock In {\em Conference Record of the Thirty-Eighth Asilomar Conference on
  Signals, Systems and Computers, 2004.}, volume~2, pages 1407--1411. IEEE,
  2004.

\bibitem{stankovic2010decentralized}
Srdjan~S Stankovi{\'c}, Milo{\v{s}}~S Stankovic, and Du{\v{s}}an~M Stipanovic.
\newblock Decentralized parameter estimation by consensus based stochastic
  approximation.
\newblock {\em IEEE Transactions on Automatic Control}, 56(3):531--543, 2010.

\bibitem{das2013Allerton}
Subhro Das and José M.~F. Moura.
\newblock Distributed linear estimationof dynamic random fileds.
\newblock In {\em 51st Annual Allerton Conference on Communication, Control,
  and Computing}, pages 1120--1125, 2013.

\bibitem{das2013Asilomar}
Subhro Das and José M.~F. Moura.
\newblock Distributed {K}alman filtering and network tracking capacity.
\newblock In {\em 47th Asilomar Conference on Signals, Systems, and Computers},
  pages 629--633, 2013.

\bibitem{lopes2008diffusion}
Cassio~G Lopes and Ali~H Sayed.
\newblock Diffusion least-mean squares over adaptive networks: Formulation and
  performance analysis.
\newblock {\em IEEE Transactions on Signal Processing}, 56(7):3122--3136, 2008.

\bibitem{sayed2013diffusion}
Ali~H Sayed, Sheng-Yuan Tu, Jianshu Chen, Xiaochuan Zhao, and Zaid~J Towfic.
\newblock Diffusion strategies for adaptation and learning over networks: an
  examination of distributed strategies and network behavior.
\newblock {\em IEEE Signal Processing Magazine}, 30(3):155--171, 2013.

\bibitem{rad2010distributed}
Kamiar~Rahnama Rad and Alireza Tahbaz-Salehi.
\newblock Distributed parameter estimation in networks.
\newblock In {\em 49th IEEE Conference on Decision and Control (CDC)}, pages
  5050--5055. IEEE, 2010.

\bibitem{bianchi2011convergence}
Pascal Bianchi, Gersende Fort, Walid Hachem, and J{\'e}r{\'e}mie Jakubowicz.
\newblock Convergence of a distributed parameter estimator for sensor networks
  with local averaging of the estimates.
\newblock In {\em 2011 IEEE International Conference on Acoustics, Speech and
  Signal Processing (ICASSP)}, pages 3764--3767. IEEE, 2011.

\bibitem{chung1997spectral}
Fan~RK Chung.
\newblock {\em Spectral Graph Teory}, volume~92.
\newblock American Mathematical Society, 1997.

\bibitem{das2022observability}
Subhro Das.
\newblock On observability and optimal gain design for distributed linear
  filtering and prediction.
\newblock {\em arXiv preprint arXiv:2203.03521}, 2022.

\bibitem{das2013ICASSP}
Subhro Das and José M.~F. Moura.
\newblock Distributed state estimation in multi-agent networks.
\newblock In {\em 38th IEEE International Conference on Acoustics, Speech and
  Signal Processing}, pages 4246--4250, 2013.

\bibitem{das2013EUSIPCO}
Subhro Das and José M.~F. Moura.
\newblock Distributed {K}alman filtering.
\newblock In {\em 21st European Signal Processing Conference}, pages 1--5,
  2013.

\bibitem{das2015TSP}
Subhro Das and José M.~F. Moura.
\newblock Distributed {K}alman filtering with dynamic observations consensus.
\newblock {\em IEEE Transactions on Signal Processing}, 63(17):4458--4473,
  2015.

\bibitem{das2017TSP}
Subhro Das and José M.~F. Moura.
\newblock Consensus+innovations distributed {K}alman filter with optimized
  gains.
\newblock {\em IEEE Transactions on Signal Processing}, 65(2):467--481, 2017.

\bibitem{sebastian2021all}
Eduardo Sebasti{\'a}n, Eduardo Montijano, and Carlos Sag{\"u}{\'e}s.
\newblock All-in-one: Certifiable optimal distributed kalman filter under
  unknown correlations.
\newblock {\em arXiv preprint arXiv:2105.15061}, 2021.

\end{thebibliography}

\end{document}